\documentclass[twocolumn,showpacs,preprintnumbers,amsmath,amssymb,prb]{revtex4}

\usepackage{graphicx}
\usepackage{dcolumn}
\usepackage{bm}

\begin{document}

\preprint{}

\title{Effect of nuclear spins on the electron spin dynamics %
in negatively charged InP quantum dots.}

\author{Ivan~V.~Ignatiev}
\email{ivan_ignatiev@mail.ru} \affiliation{Institute of Physics,
University of Tsukuba, Tsukuba 305--8571, Japan}
\affiliation{Institute of Physics, St.-Petersburg State
University, St.-Petersburg, 198504, Russia}

\author{Il'ya~Ya.~Gerlovin}
\affiliation{Vavilov State Optical Institute, St.-Petersburg, 190034, Russia}

\author{Sergey~Yu.~Verbin}
\affiliation{Institute of Physics, University of Tsukuba, Tsukuba
305--8571, Japan} \affiliation{Institute of Physics,
St.-Petersburg State University, St.-Petersburg, 198504, Russia}

\author{W.~Maruyama}
\affiliation{Institute of Physics, University of Tsukuba, Tsukuba
305--8571, Japan}

\author{Bipul~Pal}
\affiliation{Institute of Physics, University of Tsukuba, Tsukuba
305--8571, Japan}

\author{Yasuaki~Masumoto}
\affiliation{Institute of Physics, University of Tsukuba, Tsukuba
305--8571, Japan}

\date{\today}

\begin{abstract}
Kinetics of polarized photoluminescence of the negatively
charged InP quantum dots in weak magnetic field is studied
experimentally. Effect of both the nuclear spin fluctuations and
the dynamical nuclear polarization on the electron spin
orientation is observed.
\end{abstract}

\pacs{78.47.+p, 72.25.Fe, 73.21.La}

\maketitle

\section*{Introduction}
Strong localization of electrons in quantum dots (QDs) may
considerably enhance hyperfine interaction of electron spins with
those of nuclei \cite{1}. Two effects of the interaction are
possible.  First, due to limited number of nuclear spins
interacting with the electron spin in a QD, typically $n \sim
10^5$, random correlation of nuclear spins may create a
fluctuating nuclear polarization, $\Delta S_N \propto
S_N/\sqrt{n}$, where $S_N$ is the total spin of the polarized
nuclei. Fluctuation $\Delta S_N$ acts on the electron spin as an
effective magnetic field, $\delta B_{N}$, with random magnitude
and orientation \cite{2}. Electron spin precession in this field
results in the relatively fast dephasing of the electron spins in
ensemble of QDs and in the three-fold decrease in magnitude of the
total electron spin. Theoretical estimates for the GaAs QDs give
rise to sub-nanosecond dephasing times \cite{2}. This dephasing
may prohibit from realization of the spin memory devices proposed
by the so-called "Spintronics" \cite{3}. The fluctuating nuclear
polarization may also hinder from study of the electron spin
relaxation in steady-state conditions by means of Hanle effect
\cite{2,4}. Problem is that the electron spin does not "feel" any
external magnetic field while the field is weaker than the
effective magnetic field of the fluctuations \cite{2}.

Second effect of the hyperfine interaction appears when the
nuclear spins are polarized, e.g., by means of optical orientation
of the electron spin in  presence of external magnetic field
\cite{5}. The dynamic nuclear polarization may act as a relatively
strong effective magnetic field, $B_{N}$, causing Zeeman splitting
of the electronic sub-levels \cite{1}. Thermalization of the
electron spin to the lowest sub-level at the low temperature
(freezing effect) may create the electron spin polarization which
lives as long as the nuclear spin polarization lives. In this
case, the long-lived spin memory can be realized due to the
freezing effect rather than to slow electron spin relaxation
theoretically justified for QDs by Khaetskii and Nazarov~\cite{6}.

In this work, we present the experimental study of nuclear field
effects on long-lived spin polarization in singly charged InP QDs
observed recently \cite{11,Pal}.

\section {Experimental}
We have studied a sample with the single layer of the QDs grown
between the InGaP barrier layers on the n-doped GaAs substrate by
the gas source MBE technology. Semi-transparent indium-tin-oxide
electrode was deposited on top of the sample to control the
charged state of the dots by means of applied bias. It was found
previously \cite{10} that the QDs contain one resident electron
per dot (in average) at $U_{bias} = -0.1$~V.

To polarize spins of the resident electrons, we have used an
excitation by the circularly polarized light at the energy which
is slightly larger ($\sim 40$~meV) than the energy of the lowest
optical transition in the QDs (intra-dot excitation). Such
excitation creates an electron-hole pair at an excited state.
After energy relaxation followed by electron-hole recombination,
spin orientation of the pair can be transferred to spin of the
resident electron and conserved for a time interval much longer
than the recombination time. Mechanisms of the spin orientation
are widely discussed in literature (see, e.g., \cite{5}).

\section{Experimental results and discussion.}
For the QDs under study, spin polarization of the resident
electrons can be detected by means of negative circular
polarization of photoluminescence (PL) observed in the PL spectrum
and kinetics \cite{11}. An example of the polarization kinetics is
shown in Fig~1. As it is seen, the polarization degree approaches
some constant negative value called hereafter as the amplitude of
the negative circular polarization (NCP) of the PL. Mechanism of
the NCP formation is discussed in Ref.~\cite{11} where it is shown
that the NCP amplitude reflects spin orientation of the resident
electrons.

\begin{figure}[hbt]
\includegraphics*[width=8cm]{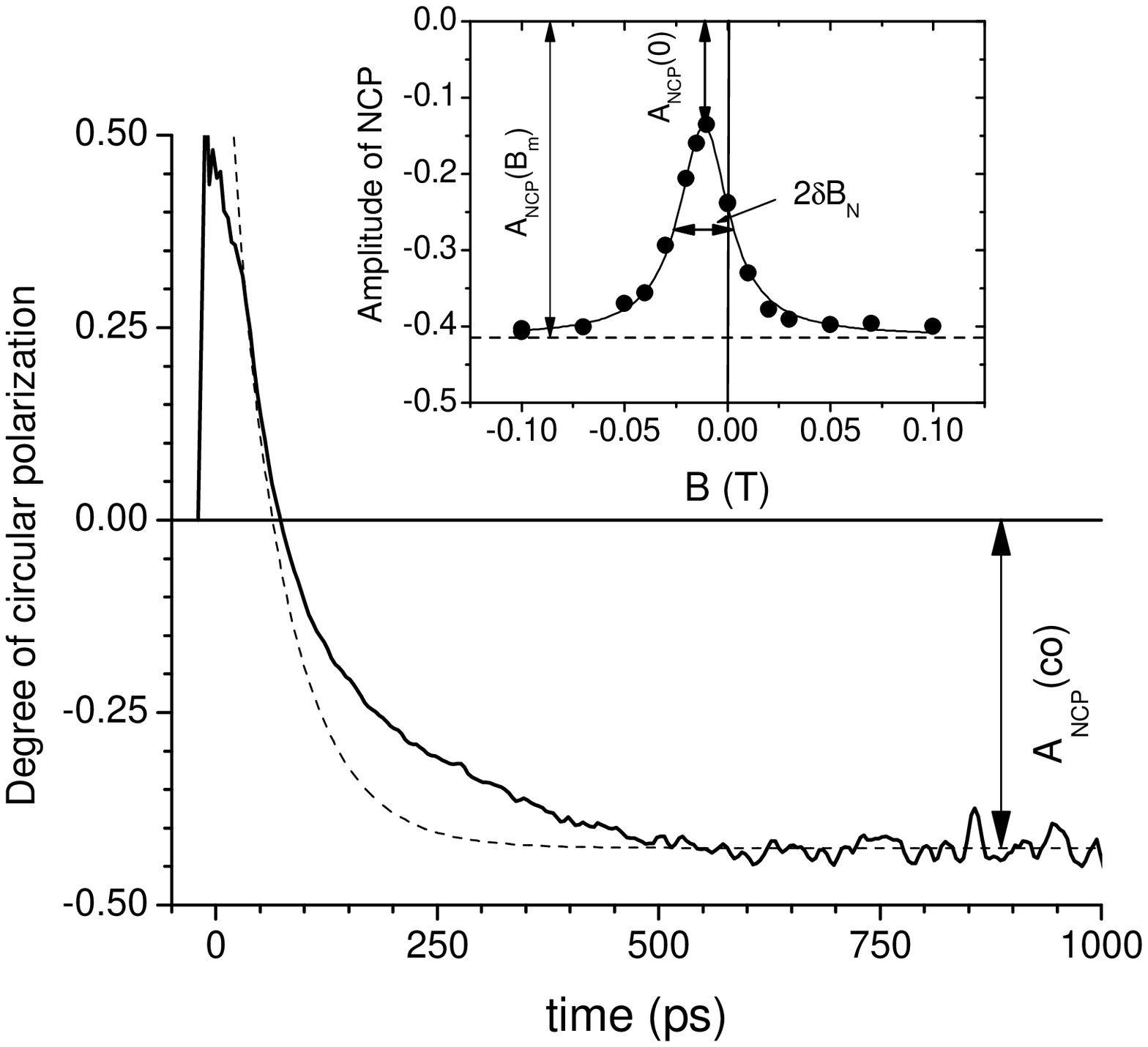}
\caption{Kinetics of degree of circular polarization. Sample
temperature $T=5$~K. Dashed line is the fit in framework of a
model of the NCP formation. Inset: dependence of the NCP amplitude
on longitudinal magnetic field. }
\end{figure}

Inset in Fig.1 shows dependence of the NCP amplitude on the
relatively small magnetic field applied along the optical axis of
the excitation (Faraday configuration). The curve has a pronounced
maximum at $B \sim 0$ which is well fitted by a Lorentzian curve
with the full width at half maximum (FWHM), $\Delta =30$~mT. The
NCP amplitude at moderate magnetic field ($B>0.1$~T) is three
times larger in absolute value than that in maximum of the
dependence. Similar behavior of circular polarization of PL was
observed in Refs.~\cite{4,7,8,9}.

In accordance with the theory \cite{2}, three-fold decrease in the
NCP amplitude at $B \sim 0$ can be caused by the fluctuating
nuclear polarization. For simple explanation of the of the NCP
decrease, random orientation of the nuclear spin fluctuations can
be replaced by the three types of the fluctuations directed along
$x$, $y$, and $z$ axes with equal probabilities. Electron spins
directed along $z$ axis will be dephased due to precession in
nuclear fields, $\delta B_N$, directed along $x$ and $y$ axes but
partly conserved for $z$ component of the field.

In presence of the longitudinal ($z$-oriented) external magnetic
field, $B_{ext}$, electron spin "feels" the total field which is
the vector sum of $\delta B_N$ and $B_{ext}$. Therefore effect of
the nuclear spin fluctuations becomes negligible with the
$B_{ext}$ rise. So the dependence $A_{NCP}(B_{ext})$ must reveal a
singularity near zero value of $B_{ext}$.  FWHM of the singularity
allows us to estimate the strength of $\delta B_N$. As it is shown
in the inset in Fig. 1, in our case this value is about 15 mT.

\begin{figure}[hbt]
\includegraphics*[width=8cm]{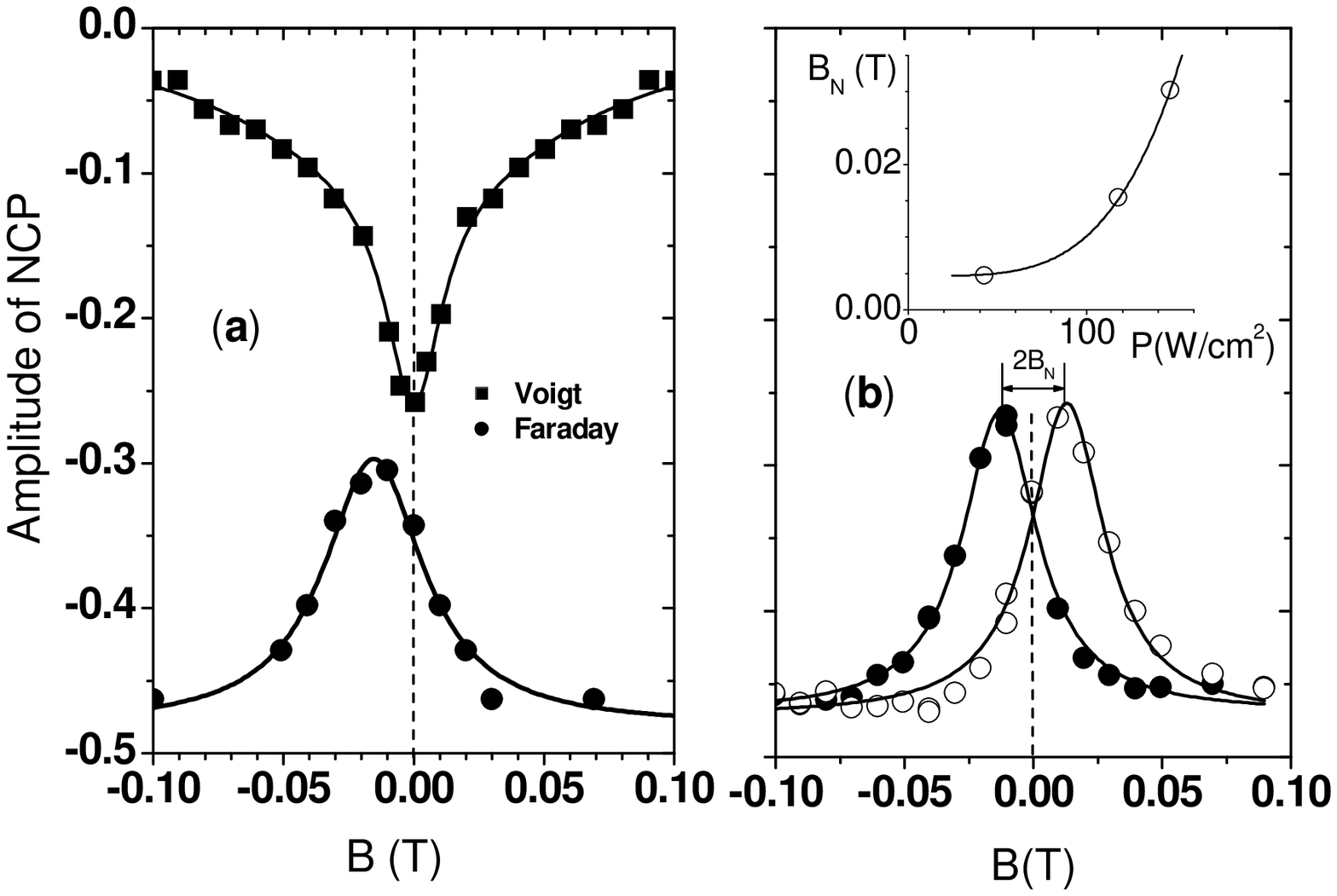}
\caption{(a) Dependence of the NCP amplitude on the longitudinal
(circles) and transverse (squares) magnetic field. Solid lines are
fits by Lorentzians. (b) Magnetic field dependence of the NCP
amplitude measured at $\sigma^+$- (filled circles) and
$\sigma^-$-(empty circles) polarized excitations. Inset shows
power dependence of the effective nuclear magnetic field.}
\end{figure}

In Fig.~2(a), the dependence of the NCP amplitude on the magnetic
field in Faraday configuration ($B \parallel z$) is compared with
that in Voigt configuration ($B \perp z$). In the latter case, the
NCP amplitude decreases in the magnetic field due to the electron
spin precession around the magnetic field direction (Hanle
effect). According to classical theory of the Hanle effect
\cite{5}, FWHM of the magnetic field dependence, $\Delta B_H$,
should be related to the spin relaxation time, $\tau_s$:
$$\tau_s = \hbar / \left(g \mu_B \Delta B_H\right),$$
where $g$ and $\mu_B$ are the electron $g$-factor and Bohr
magneton respectively. Substituting experimentally determined
value $\Delta B_H = 10$~mT to the above equation results in
unreasonably small value of the spin relaxation time $\tau_s
\approx 1$~ns which is approximately five orders of magnitude
smaller than that measured in the direct experiments \cite
{11,Pal}.

At the same time, the FWHM of the magnetic field dependence in
Voigt configuration is very close to that in Faraday configuration
[see Fig. 2(a)]. So, it is natural to suppose that, in our case,
the Hanle effect is the result of competition between the external
magnetic field and the fluctuations of nuclear field in the QDs.

In presence of the external longitudinal magnetic field and
optical excitation, dynamic nuclear polarization may occur as it
is discussed in the introduction. To estimate the effective
magnetic field created by dynamical nuclear polarization, $B_N$,
we have measured field dependencies of the NCP amplitude in
Faraday configuration for two opposite circular polarization of
the exciting light. The results of the experiment are shown in
Fig. 2(b). The relative shift of the curves corresponds to the
value of the dynamical nuclear field $B_N = 15$~mT. This value is
the same order of magnitude as that obtained in Refs.~\cite{7,9}.
As it is seen from inset in Fig. 2(b), the value of $B_N$
increases with increasing of the laser beam power, i.e., the
nuclear field is really created by optical pumping. It should be
emphasized that the power dependence of $B_N$ is clearly
superlinear which is an indication of the threshold nature of the
dynamical nuclear orientation process. At the same time, the
values of $B_N$ optically created in InP QDs are much smaller than
those for GaAs QDs presented in Ref.~\cite{1}. Further studies are
needed to understand this strong distinction between two types of
QDs.

\section {Conclusion}
In conclusion, we have studied experimentally the effect of the
small magnetic fields on the electron spin orientation in InP QDs.
The abrupt increase of the spin orientation was found in the
longitudinal fields larger than 0.015 T. In the transverse fields,
the spin orientation abruptly decreased at the same values of the
field. This behavior was explained as a result of competition
between fluctuations of nuclear field in the QD and the external
magnetic field. Mean value of the fluctuations of the nuclear
field and the strength of the effective magnetic field created by
the dynamical nuclear polarization are estimated from the
experiments to be few tens of meV.

\section*{Acknowledgments}
{This work was supported by Grant-in-Aid for Scientific Research
No. 13852003 and No. 16031203 from the MEXT of Japan, by the
Russian Foundation for Basic Research (project No. 0302-16858), by
INTAS (project No. 1B 2167), and by ISTC (project No.~2679).}

\end{document}